# Relations between near-field enhancements and Purcell factors in hybrid nanostructures of plasmonic antennas and dielectric cavities

Xu-Tao Tang, Lin Ma, Yue You, Xiao-Jing Du, Hua Qiu, Xi-Hua Guan, Jun He, and Zhong-Jian Yang*

*Hunan Key Laboratory of Nanophotonics and Devices, School of Physics, Central South University, Changsha 410083, China*

*E-mail: zjyang@csu.edu.cn

**Abstract:** Strong near-field enhancements (NFEs) of nanophotonic structures are believed to be closely related to high Purcell factors ($F_P$). Here, we theoretically show that the correlation is partially correct; the extinction cross section ($\sigma$) response is also critical in determining $F_P$. The divergence between NFE and $F_P$ is especially pronounced in plasmonic-dielectric hybrid systems, where the plasmonic antenna supports dipolar plasmon modes and the dielectric cavity hosts Mie-like resonances. The cavity's enhanced-field environment can boost the antenna's NFEs, but the $F_P$ is not increased concurrently due to the larger effective $\sigma$ that is intrinsic to the $F_P$ calculations. Interestingly, the peak $F_P$ for the coupled system can be predicted by using the NFE and $\sigma$ responses. Furthermore, the limits for $F_P$ of coupled systems are considered; they are determined by the *sum* of the $F_P$ of a redshifted (or modified, if applicable) antenna and an individual cavity. This contrasts starkly with the behavior of NFE which is closely associated with the *multiplicative* effects of the NFEs provided by the antenna and the dielectric cavity. The differing behaviors of NFE and

$F_P$ in hybrid cavities have varied impacts on relevant nanophotonic applications such as fluorescence, Raman scattering and enhanced light-matter interactions.

## 1. Introduction

Plasmonic antennas can concentrate incident light into deeply subwavelength regions, known as hot spots, due to strong surface plasmon resonances. The near-field enhancements (NFEs) in these hot spot regions can reach up to ~$10^2$ or even higher [1, 2]. Subwavelength dielectric cavities, made of high-refractive index materials, can also exhibit strong far-field optical responses with considerable near-field enhancements due to their Mie-like resonances [3, 4]. The near field enhancements of common resonances in dielectric cavities are usually lower compared to their plasmonic counterparts. Nonetheless, the NFEs of certain special modes in subwavelength dielectric cavities, such as whispering gallery modes (WGMs), can still be quite high [5]. The significant NFEs in those photonic structures are usually accompanied by highly enhanced Purcell factors ($F_p$), which refer to the enhancements in the spontaneous decay rates of an emitter. [5-14]. These near-field properties have enabled a wide range of applications such as enhanced fluorescence, Raman scattering, nanolasers, enhanced light-matter interactions and so on [15-24].

The enhanced near field properties of plasmonic antennas are associated with their highly concentrated field and small sizes, namely, small mode volumes ($V$) [6, 10, 25-28]. In contrast, the enhanced near field properties of dielectric cavities are achieved by resonances of high enough quality ($Q$) factors [5, 6]. The $F_p$ of a plasmonic or dielectric nanostructure can be expressed in terms of these factors as $F_p \propto Q/V$ [6]. However, the $Q$ factors of common plasmonic antennas are usually low (~$10^1$), and the field concentrations of common dielectric cavities are low (large mode volumes), meaning the fields are widely spread inside the whole cavities [4]. Recently, hybrid photonic cavities of plasmonic antennas and dielectric cavities have been drawing increasing interests [29-33]. These hybrids may offer the potential to combine the advantages of each component. A dielectric cavity can readily provide enhanced field environment for a small plasmonic antenna due to their differing geometric sizes. This arrangement has enabled increased NFEs around plasmonic antennas [29, 34-39]. The relative increase in NFE for a plasmonic antenna can reach more than ~$10^1$. $F_p$ responses have also been studied in terms of large values, spectral modifications, and directional emissions [36, 40-45]. Referring to the reports on NFE and $F_P$ in individual dielectric or plasmonic nanostructures, it seems intuitive that a larger NFE would indicate a larger Fp. As for hybrid systems, the relation between NFE and $F_P$ has been scarcely investigated.

Here, we theoretically show that enlarged NFEs in nanophotonic cavities do not necessarily indicate enlarged $F_P$. This is particularly evident in plasmonic-dielectric hybrid cavities. We consider a general plasmonic-dielectric hybrid cavity through a semi-analytical model. A small plasmonic antenna exhibits a dipolar surface plasmon resonance, while a large dielectric cavity has a Mie-like resonance. The antenna is assumed to be exposed to an effective field with a certain value provided by the dielectric cavity. The results based on analytical model and direct numerical simulations agree well. The NFE of a hybrid cavity ($|E/E_0|_{\text{tot}}$) is closely related to the *product* of NFEs provided by the antenna and dielectric cavity. However, the enlarged NFE of the antenna does not lead to a simultaneous increase in the $F_\text{P}$ of a hybrid cavity. This is due to the fact that extinction cross section ($\sigma$) also plays a crucial role in determining the $F_\text{P}$. Interestingly, the peak $F_\text{P}$ of coupled system can be estimated by the NFE and $\sigma$ responses. Both the $|E/E_0|_{\text{tot}}$ and $F_\text{P}$ values also depend on the antenna-cavity coupling strength. Nonetheless, the upper limit for $F_\text{P}$ of a hybrid cavity is set by the *sum* of $F_\text{P}$ provided by individual cavity and redshifted antenna (or modified antenna, if applicable). These findings are important for nanophotonic applications such as enhanced fluoresces, Raman scattering and strong light-matter couplings. These relevant impacts will also be discussed.

## 2. Theoretical model

We consider a general hybrid cavity of a plasmonic antenna and a dielectric cavity (Figure 1). The antenna exhibits a common dipolar surface plasmon resonance, while the dielectric cavity has a Mie-like resonance. The whole hybrid structure is excited by a plane wave or a dipole emitter. The NFE of the hybrid system can be obtained by considering a plane wave as the excitation (Figure 1a). The total field felt by the antenna can be expressed as $E_a = E_0 + XP_c$ [46], where $E_0$ is the electric field of the excitation plane wave, $XP_c$ is the contribution from the dielectric cavity, $P_c$ is the moment of an electromagnetic mode, and $X$ represents the proportional coefficient between the electric field and the moment $P_c$ at the location of the antenna. The total field felt by the cavity can be written as $E_c = E_0 + XP_a$, where $P_a$ is the dipole moment of the antenna, namely, $P_a = \varepsilon_0 \alpha_a E_a$. $\alpha_a$ is the electric dipolar polarizability of the antenna, and $\varepsilon_0$ is the free space permittivity. Similarly, $P_c$ can be written as $P_c = \varepsilon_0 \alpha_c E_c$, where $\alpha_c$ is the polarizability of the cavity. Then, one obtains

$$\begin{cases} P_a = \frac{\varepsilon_0 \alpha_a (1 + X\varepsilon_0 \alpha_c)}{1 - X^2 \varepsilon_0^2 \alpha_a \alpha_c} E_0 \\ P_c = \frac{\varepsilon_0 \alpha_c (1 + X\varepsilon_0 \alpha_a)}{1 - X^2 \varepsilon_0^2 \alpha_a \alpha_c} E_0 \end{cases}.$$

Now, if one considers the field of a point L in the hybrid (Figure 1a). It can be expressed as $E = E_0 + YP_a + ZP_c$, where $Y$ ($Z$) represents the proportional coefficient between the electric field at point L and the moment of antenna (cavity). Thus, the total NFE $\left|\frac{E}{E_0}\right|_{tot}$ at a point L can be expressed as

$$\left|\frac{E}{E_0}\right|_{tot} = \left|\frac{\varepsilon_0 (Y\alpha_a + Z\alpha_c + XY\varepsilon_0 \alpha_a \alpha_c + XZ\varepsilon_0 \alpha_a \alpha_c)}{1 - X^2 \varepsilon_0^2 \alpha_a \alpha_c} + 1\right|. \qquad (1)$$

If the direct contribution from the cavity to the NFE at L can be ignored ($Z \approx 0$), Eq. (1) reduces to

$$\left|\frac{E}{E_0}\right|_{tot} = \left|\frac{\varepsilon_0 (Y\alpha_a + XY\varepsilon_0 \alpha_a \alpha_c)}{1 - X^2 \varepsilon_0^2 \alpha_a \alpha_c} + 1\right|. \qquad (2)$$

For the $F_P$, the hybrid system is excited by an electric dipole emitter (Figure 1b). The emitter is located at the same point L. The total field felt by the antenna can be expressed as $E_a = YD + XP_c$, where $D$ is the electric dipole moment of the emitter. The other parameters are the same as before. The total field felt by the cavity can be expressed as $E_c = XP_a + ZD$. Then, one obtains

$$\begin{cases} P_a = \frac{\varepsilon_0 \alpha_a (Y + XZ\varepsilon_0 \alpha_c)}{1 - X^2 \varepsilon_0^2 \alpha_a \alpha_c} D \\ P_c = \frac{\varepsilon_0 \alpha_c (Z + XY\varepsilon_0 \alpha_a)}{1 - X^2 \varepsilon_0^2 \alpha_a \alpha_c} D \end{cases}.$$

The total field felt by the quantum emitter $E'$ can be expressed as $E' = YP_a + ZP_c$. So, the power radiated from the emitter can be written as

$$P = \frac{\omega_0}{2} \text{Im}(DE') = \frac{\omega_0}{2} \text{Im}\left(\frac{Y^2 \varepsilon_0 \alpha_a + 2XYZ\varepsilon_0^2 \alpha_a \alpha_c + Z^2 \varepsilon_0 \alpha_c}{1 - X^2 \varepsilon_0^2 \alpha_a \alpha_c} D^2\right), \qquad (3)$$

Here, we have assumed that the direction of the dipole moment $D$ is the same as that of the $E'$ for simplicity. Thus, the $F_P$ of such a system (denoted by $F_{\text{P-tot}}$) is

$$F_{\text{P-tot}} = \frac{P}{P_0} = \frac{6\pi\varepsilon_0 c^3}{\omega_0^3} \text{Im}\left(\frac{Y^2 \varepsilon_0 \alpha_a + 2XYZ\varepsilon_0^2 \alpha_a \alpha_c + Z^2 \varepsilon_0 \alpha_c}{1 - X^2 \varepsilon_0^2 \alpha_a \alpha_c}\right), \qquad (4)$$

where $P_0$ is the power radiated from the emitter in vacuum $P_0 = \frac{\omega_0^4}{12\pi\varepsilon_0 c^3} D^2$, $\omega_0$ and $c$ are the radiation frequency and speed of light, respectively. If the direct contribution from the cavity can also be ignored ($Z \approx 0$), $F_{\text{P-tot}}$ reduces to

$$F_{P-tot} = \frac{6\pi\varepsilon_0 c^3}{\omega_0^3} \text{Im}\left(\frac{Y^2 \varepsilon_0 \alpha_a}{1 - X^2 \varepsilon_0^2 \alpha_a \alpha_c}\right). \qquad (5)$$

For the sake of later discussion, Eq. (5) can be divided into the coupling part $\text{Im}\left(\frac{Y^2 \varepsilon_0 \alpha_a}{1 - X^2 \varepsilon_0^2 \alpha_a \alpha_c}\right)$ and frequency part $\frac{6\pi\varepsilon_0 c^3}{\omega_0^3}$. We define the term $X^2 \varepsilon_0^2 \alpha_a \alpha_c$ as the coupling parameter, and its value will largely affect the behaviors of the coupled system. If the coupling between antenna and cavity disappears, namely, $X = 0$. $F_{\text{P-tot}}$ reduces to the $F_P$ of an individual antenna (denoted by $F_{\text{P-A}}$)

$$F_{P-A} = \frac{6\pi\varepsilon_0 c^3}{\omega_0^3} \text{Im}(Y^2 \varepsilon_0 \alpha_a), \qquad (6)$$

which is consistent with previous reports [47]. The extinction cross section of an individual antenna $\sigma_A$ satisfies $\sigma_A = k \, \text{Im}(\alpha_a)$, where $k$ is the wave vector of the plane wave. At resonance, the $\text{Im}(E/E_0)$ can be approximately expressed as $|E/E_0|$ (denoted by $|E/E_0|_A$) [47]. Thus, the peak $F_P$ of an individual antenna (denoted by

$F_{P-A}^p$) can be rewritten as

$$F_{P-A}^p = \frac{6\pi c^2}{\omega_0^2} \frac{|E/E_0|_A^2}{\sigma_A}. \qquad (7)$$

The above parameters $X$, $Y$, $Z$, $\alpha_a$, and $\alpha_c$ are highly dependent on the geometries of antennas and cavities. Generally, they cannot be obtained by analytical solutions. Thus, we use numerical simulations to calculate these parameters. The numerical simulations are carried out by a finite-difference in time-domain (FDTD solutions) method (see the Methods section). We employ the multipole decomposition method (see the Methods section) [48] to numerically obtain $\alpha_a$ and $\alpha_c$. The near fields $XP_c$, $YP_a$, and $ZP_c$ can be directly obtained by the FDTD simulations. Then, one can obtain $X$, $Y$, and $Z$. The extinction cross sections can also be calculated straightforwardly via the FDTD simulations.

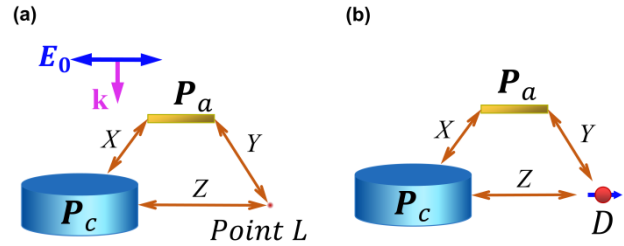

**Figure 1**. (a) Schematic of a general antenna-dielectric hybrid structure excited by a plane wave. The point L denotes the location where the NFE is investigated. (b) The same hybrid structure excited by a dipole emitter, which is located at point L.

## 3. Results and discussion

### 3.1. Relations between near-field enhancements and Purcell factors

Let us now consider a hybrid system consisting of a gold nanorod (antenna) and a silicon (Si) disk (Figure 2). The nanorod has a broad dipolar plasmon response around the considered spectral region. The disk has a narrow WGM of $m = 5$ at $\lambda = 2403$ nm

(see Figure S1) [5]. The surface distance between antenna and disk is $d = 32$ nm. We consider the total NFE ($|E/E_0|_{\text{tot}}$) around the end of antenna at the location L (Figure 2a). Here, the direct contribution from WGM to the $|E/E_0|_{\text{tot}}$ at the position L is only ~5, and it can be approximately ignored ($Z \approx 0$). The directly simulated NFE agrees well with the analytical result based on Eq. (2) (Figures 2b and 2c). The spectrum shows a Fano lineshap due to the Fano resonance of the system [41, 49, 50]. The peak value of $|E/E_0|_{\text{tot}}$ is about 8 times of the antenna at this position ($|E/E_0|_A \approx 6$). The relative enhancement is smaller than that provided by the disk cavity (~20 times), which is attributed to the broadened response brought by a relatively large coupling effect. Similar coupling effect has been reported in other nanophotonic structures [51]. The NFEs, which are away from the WGM position, obtained based on direct simulations and analytical model are slightly different. This is due to the fact that there are contributions from other modes in simulations (see Figure S1) that are not included in the analytical mode.

Let us turn to the $F_P$ of the hybrid system $F_{\text{P-tot}}$, where an emitter is taken as the excitation source (Figure 2d). The emitter is located at the same point L with its polarization aligned along the antenna. The simulated $F_{\text{P-tot}}$ also agrees well with the analytical result based on Eq. (5) (Figures 2e and 2f). Here, the direct contribution from disk cavity to the $F_{\text{P-tot}}$ is ignorable ($Z \approx 0$, see Figure S1). The $F_{\text{P-tot}}$ spectral lineshape is similar to that of the above NFE response. On the other hand, the peak $F_{\text{P-tot}}$ (denoted by $F_{\text{P-tot}}^p$) is only less than 3 times of the antenna ($F_{\text{P-A}} \approx 750$). This enhancement value is far away from the relative field-intensity enhancement at L, namely $|E/E_0|_{\text{tot}}^2/|E/E_0|_A^2 \approx 64$.

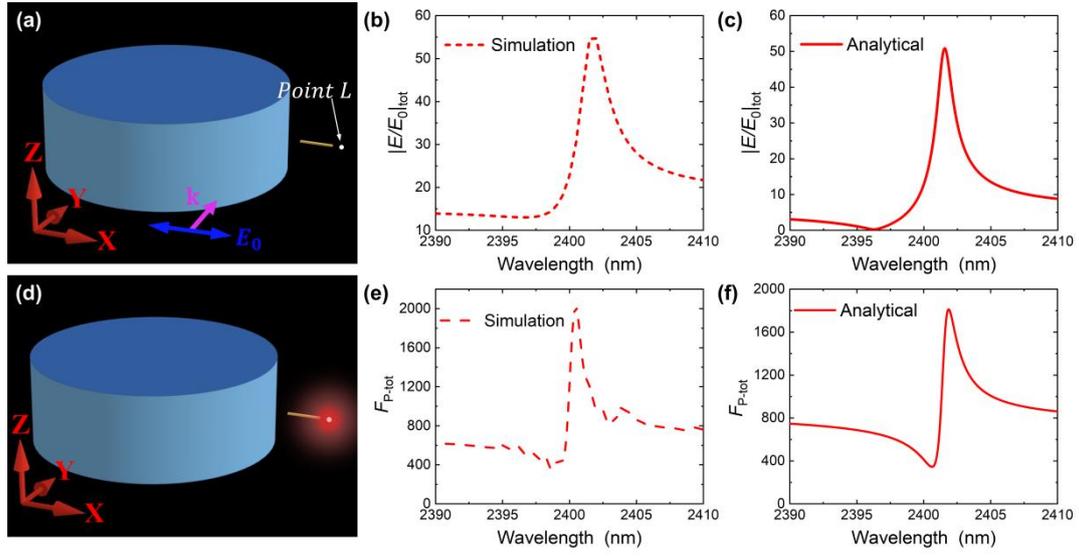

**Figure 2.** NFE and $F_P$ of a hybrid structure. The diameter and height of the disk are 2020 and 730 nm, respectively. The length, width and height of the gold nanorod are 246, 10 and 10 nm, respectively. The surface distance between disk and nanorod is 32 nm. (a) Schematic of the hybrid structure excited by a plane wave. The point L denotes a location where the NFE is investigated, and it is 21 nm away from the rod end. (b,c) The NFE at L calculated based on direct simulations (b) and the semi-analytical model (c). (d) Schematic of the same hybrid structure excited by a dipole emitter. The emitter is located at the same point of L. (e,f) The $F_P$ calculated based on direct simulations (e) and the semi-analytical model (f).

Figure 3 shows more results about the peak values of NFE and $F_{\text{P-tot}}$ as a function of the coupling parameter $|X^2\varepsilon_0^2\alpha_a\alpha_c|^{1/2}$. Here, the variation of coupling parameter is carried out by changing the NFE provided by the WGM $|E/E_0|_c$ while the other parameters remain that same as that of Figure 2. Thus, the $X$ varies correspondingly. This can be done by narrowing the gap between antenna and cavity in a realistic system. We still ignore $Z$ for simplicity. The behavior of $F_P$ clearly does not follow that of the field intensity in hybrid structures (Figures 3a and 3c). This comes from the fact that the extinction cross section $\sigma$ should also be considered for $F_P$ in addition to

NFE. Note that even within an individual component, the relation between the NFE and $F_P$ ($F_P \propto \text{NFE}^2$) holds under the default condition that the extinction $\sigma$ of the antenna is fixed. This could correspond to a common situation that an emitter moves around a given antenna (see Figure S1). The full relation in an antenna is Eq. (7), namely, $F_P \propto \text{NFE}^2/\sigma$. Thus, the relation $F_P \propto \text{NFE}^2$ without considering the $\sigma$ is only partially correct. This can be verified by comparing the cases of individual antenna and cavity (see Figure S1). As for a hybrid structure, the NFE can be enlarged significantly while the $\sigma$ of the coupled system is also much larger than that of the individual antenna (Figures 3b and S1).

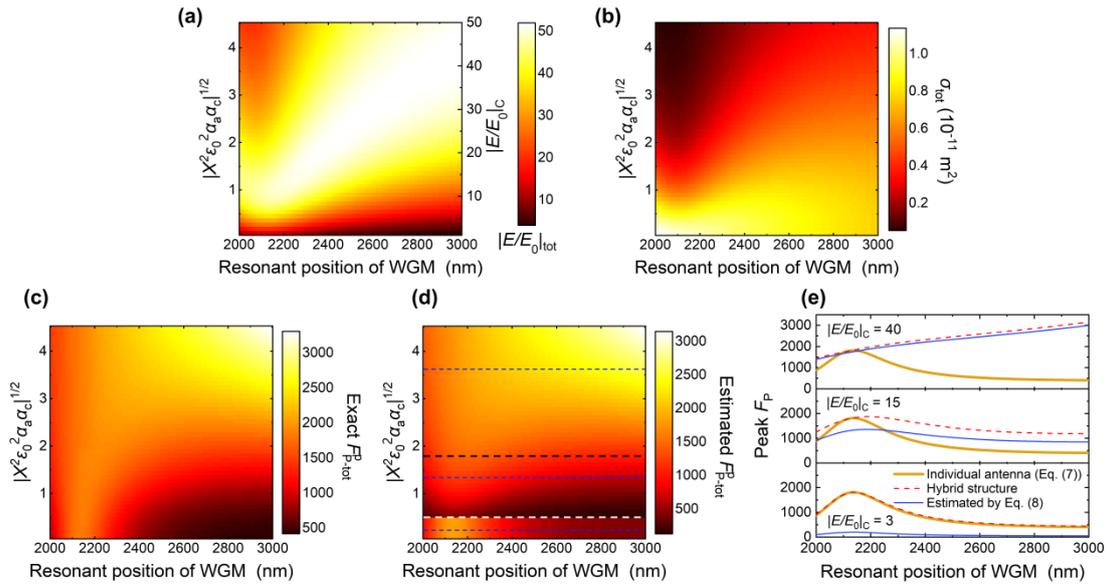

**Figure 3.** Relations between NFE and $F_P$ of hybrid structures. (a-c) The peak NFE (a), $\sigma$ (b) and $F_{\text{P-tot}}$ (c) of the hybrid system as a function of the coupling parameter $|X^2\varepsilon_0^2\alpha_a\alpha_c|^{1/2}$ and resonant position of WGM. Here, the variation of coupling parameter $|X^2\varepsilon_0^2\alpha_a\alpha_c|^{1/2}$ is carried out by changing the NFE provided by the WGM ($|E/E_0|_c$) as demonstrated by the right axis in (a). (d) The estimated $F_{\text{P-tot}}^{\text{p}}$ as a function of $|X^2\varepsilon_0^2\alpha_a\alpha_c|^{1/2}$ and resonant position of WGM. The results above and below the dotted white line are based on Eq. (8) and (7), respectively. (e) The estimated $F_{\text{P-tot}}^{\text{p}}$ with three different $|X^2\varepsilon_0^2\alpha_a\alpha_c|^{1/2}$ ($|E/E_0|_c$). The exact $F_{\text{P-tot}}^{\text{p}}$ are also shown for comparison. The data are also marked by the corresponding dotted blue lines in (d).

The peak $F_{\text{P-tot}}$ (= $F^{\text{p}}_{\text{P-tot}}$) in a hybrid system can be estimated by the peak NFE and the $\sigma$ responses. Let us first consider the case of a weak enough coupling where the term $|X^2\varepsilon_0{}^2\alpha_a\alpha_c| \ll 1$. Here, we still ignore the direct contribution from the cavity for simplicity ($Z \approx 0$). Under such a condition, Eq. (5) reduces to Eq. (6), namely, $F^{\text{p}}_{\text{P-tot}} \approx F^{\text{p}}_{\text{P-A}}$. This indicates that $F^{\text{p}}_{\text{P-tot}}$ is largely determined by that of the uncoupled antenna under weak enough coupling. Interestingly, for the case with strong enough coupling where the coupling parameter satisfies $|X^2\varepsilon_0{}^2\alpha_a\alpha_c| \gtrsim 3$, the $F^{\text{p}}_{\text{P-tot}}$ can be well estimated by

$$F^{\text{p}}_{\text{P-tot}} = \frac{6\pi c^2}{\omega_0{}^2} \frac{|E/E_0|_{tot}{}^2}{\sigma_{tot}}, \qquad (8)$$

where $\sigma_{tot}$ is the extinction cross section of the hybrid system. This expression is similar to Eq. (7), and one only needs to replace the NFE and $\sigma$ of an individual antenna by those of the hybrid structure, respectively.

Figures 3c and 3d shows the exact and estimated results of $F^{\text{p}}_{\text{P-tot}}$ for a case with varying the coupling strength, respectively. For weak and strong couplings, the estimated results based on Eqs. (7) and (8) show good agreements with the exact values, respectively. It is also found that for moderate couplings ($0.3 < |X^2\varepsilon_0{}^2\alpha_a\alpha_c| < 3$), the $F^{\text{p}}_{\text{P-tot}}$ can also be roughly estimated by Eq. (8), while the value is underestimated. The estimated values based on Eq. (8) are about 30% to 80% of the exact values, corresponding to the region between the white and black lines in Figure 3d. Further calculations show that the above estimations and the corresponding accuracies also apply for cases with different $\alpha_a$ in the coupling parameter (see Figure S2). The results indicate that if one takes the hybrid system as a whole, the enlarged NFE will always be accompanied by a larger effective $\sigma$ involved for the $F_{P-\text{tot}}^p$. This effective $\sigma$ is larger than that of the uncoupled antenna.

**3.2. Limits for Purcell factors of hybrid systems with different coupling strengths**

Now let us consider the limit for $F_P$ of a hybrid system. Here, we assume a system with a given antenna ($F_{P-A}$) and study the effects of mode couplings on the total $F_P$ of the system. We are not intended to optimize the $F_P$ value of a general nanophotonic system [52, 53]. To get more insight into the limit for $F_P$ of a system with $Z \approx 0$, we shall focus on the antenna-cavity coupling part of Eq. (5), namely, $\text{Im}\left(\frac{\alpha_a}{1-X^2\varepsilon_0^2\alpha_a\alpha_c}\right)$. As the $Q$ factor of a dielectric resonance (WGM) is much higher than the plasmon resonance, the $\alpha_c$ at a wavelength far away from its resonance can be taken as $\alpha_c \approx 0$. One obtains $\text{Im}\left(\frac{\alpha_a}{1-X^2\varepsilon_0^2\alpha_a\alpha_c}\right) \approx \text{Im}(\alpha_a)$. This means that $F_P$ is determined by the antenna for wavelengths far away from the cavity mode.

For the wavelengths around cavity resonance, $\alpha_a$ can be taken as a fixed value as its variation with wavelength is relatively much slower than that of $\alpha_c$. The response of the cavity mode can be approximately taken as a Lorentz resonance. Then, $\alpha_c$ can be written as $\begin{cases}\text{Re}(\alpha_c) = D_c \sin\theta_c \cdot \cos\theta_c \\ \text{Im}(\alpha_c) = D_c \cos^2\theta_c\end{cases}$, where $\theta_c$ is a parameter that is related to wavelength and $D_c$ is the maximal value of $\text{Im}(\alpha_c)$, namely, $D_c = \text{Max}[\text{Im}(\alpha_c)]$ (see Figure S3). The peak value of $\text{Im}\left(\frac{\alpha_a}{1-X^2\varepsilon_0^2\alpha_a\alpha_c}\right)$ can be approximately written as (see SI for the relevant derivation)

$$\text{Peak}\left[\text{Im}\left(\frac{\alpha_a}{1-X^2\varepsilon_0^2\alpha_a\alpha_c}\right)\right] \approx \frac{X^2\varepsilon_0^2|\alpha_a|^2 D_c + \text{Im}(\alpha_a)}{X^2\varepsilon_0^2\text{Im}(\alpha_a)D_c + 1}. \quad (9)$$

To obtain the maximal value of the peak $\text{Im}\left(\frac{\alpha_a}{1-X^2\varepsilon_0^2\alpha_a\alpha_c}\right)$, the response of the dipolar plasmon mode can also be taken as a Lorentz lineshape. $\alpha_a$ can be written as

$$\begin{cases} \text{Re}(\alpha_a) = D_a \sin\theta_a \cdot \cos\theta_a \\ \text{Im}(\alpha_a) = D_a \cos^2\theta_a \end{cases}, \text{ where } \theta_a \text{ is a parameter that is also related to}$$

wavelength and $D_a$ is the maximal value of $\text{Im}(\alpha_a)$ ($D_a = \text{Max}[\text{Im}(\alpha_a)]$) which occurs at $\theta_a = 0$. Then, one can obtains (see SI for the relevant derivation)

$$\text{Max}\left[\frac{X^2\varepsilon_0^2|\alpha_a|^2 D_c + \text{Im}(\alpha_a)}{X^2\varepsilon_0^2\text{Im}(\alpha_a)D_c + 1}\right] = \text{Max}[\text{Im}(\alpha_a)]. \quad (10)$$

Based on Eq. (10), the peak value of coupling part reaches its maximum at the resonance of the antenna ($\theta_a = 0$). More importantly, Eq. (10) indicates that both the coupling between antenna and cavity ($X$) and the properties of a dielectric cavity ($D_c$) do not affect the maximal value of $F_P$.

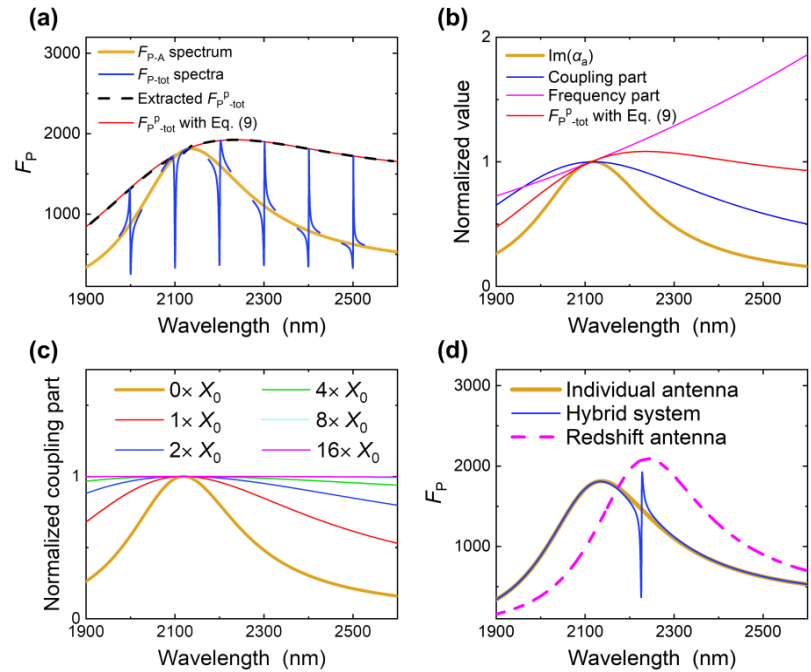

**Figure 4.** Limit for $F_P$ of a hybrid system with different couplings. The system is based on the structure in Figure 2 with $Z \approx 0$. (a) The $F_P$ spectra of hybrid structures ($F_{P\text{-tot}}$) with different WGM resonances (blue). The corresponding extracted peak value ($F_{P-\text{tot}}^p$) as a function of WGM resonance is shown by the dotted black line. The $F_{P-\text{tot}}^p$ obtained by the analytical expression Eq. (9) are plotted by the red line. The $F_P$ spectrum of an individual antenna ($F_{P\text{-A}}$) is also shown for comparison. (b) Contributions from the coupling and frequency parts to the above $F_{P-\text{tot}}^p$. The values are normalized to the peak value of $\text{Im}(\alpha_a)$, namely $D_a$. The normalized $\text{Im}(\alpha_a)$ spectrum is also shown. (c) The coupling part with varying the coupling coefficient $X$. The values are also normalized by $D_a$. The 1×$X_0$ case corresponds to the coupling strength in (a,b). (d) The $F_{\text{P-tot}}$ spectrum of a hybrid structure (blue) compared to that of a redshifted antenna (dotted orange). For the redshifted antenna, its resonance is moved to that of the WGM while the other parameters are the same as the individual one. The $F_\text{P}$ spectrum of an individual antenna is also shown.

Figure 4a shows the results of coupled systems with WGMs of different resonant positions varying around the broad response of antenna. Here, the direct contributions from the WGMs are ignorable ($Z \approx 0$). We first fix the coupling parameter $X$ for simplicity. For each case, $F_{\text{P-tot}}^\text{p}$ can be obtained based on Eq. (5), and the $F_{\text{P-tot}}^\text{p}$ as a function of the WGM position is shown (Figure 4a). The redline in Figure 4a shows the results based on the analytical expression of Eq. (9). They agree well with the results obtained from Eq. (5), which verifies the validity of the analytical formula of Eq. (9). Figure 4b shows the frequency $\frac{6\pi\varepsilon_0 c^3}{\omega_0^3}$ and coupling $\text{Im}\left(\frac{\alpha_a}{1-X^2\varepsilon_0^2\alpha_a\alpha_c}\right)$ contributions to $F_{\text{P-tot}}^\text{p}$ as the resonant position of WGM varies. The coupling part indeed cannot exceed the $D_a$ antenna, as expected (Eq. (10)). The contribution of the frequency part increases with wavelength. Thus, for $F_{\text{P-tot}}^\text{p}$ that are larger than that of an individual antenna ($F_{\text{P-A}}^\text{p}$), the contribution only comes from the frequency shift.

The coupling-part contributions with different coupling strengths are also considered (Figure 4c). With increasing the coefficient $X$, the coupling-part contributions that are spectrally distant from the antenna's resonance also increase. However, there is an upper limit which equals to $D_a$ of the antenna. This means that if the coupling between an antenna and a WGM cavity is not large enough, the coupling part will be smaller than $D_a$. Considering the frequency part, one finds that the $F_{\text{P-tot}}^\text{p}$ of hybrid system will be smaller than that of the redshifted antenna $F_{\text{P-rA}}$

(Figure 4d). Here, the $F_{\text{P-rA}}$ corresponds to an individual antenna whose resonance position is redshifted to align with that of the WGM. The other parameters of the antenna are kept the same as the original one for simplicity. Note that it is easy to obtain a redshifted antenna in a practical plasmonic system.

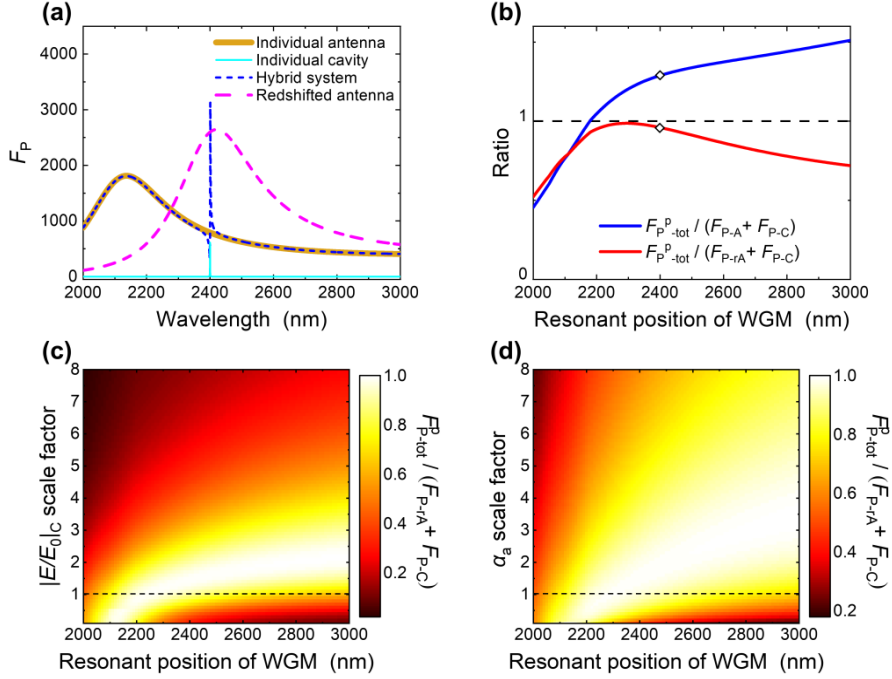

**Figure 5.** The cases with direct contributions from the WGMs ($Z \neq 0$). The other parameters are the same as that in Figure 2. (a) The $F_{\text{P-tot}}$ spectrum (blue) with a certain $F_{\text{P-C}}$ directly provided by the WGM (cyan). The other parameters are the same as that in Figure 2. The cases for the individual (yellow) and redshifted (pink) antenna are also shown. (b) Extracted peak value $F_{P-tot}^{p}$ normalized by ($F_{\text{P-rA}}$+ $F_{\text{P-C}}$) with varying the resonance of a WGM. The $F_{P-tot}^{p}$ normalized by ($F_{\text{P-A}}$+ $F_{\text{P-C}}$) are also show for comparison. The dots denote the case in (a). (c) The ratio $F_{P-tot}^{p}$ / ($F_{\text{P-rA}}$+ $F_{\text{P-C}}$) as a function of the WGM resonance and $|E/E_0|_c$. The $|E/E_0|_c$ value is normalized by that in (a,b). The dotted line corresponds to the results in (b). (d) The ratio $F_{P-tot}^{p}$ / ($F_{\text{P-rA}}$+ $F_{\text{P-C}}$) as a function of the WGM resonance and the $\alpha_a$. The $\alpha_a$ value is normalized by that in (a,b). The dotted line also corresponds to the results in (b).

For the cases with direct contributions from the WGMs, $Z$ cannot be ignored anymore. Figure 5a shows the results with a certain $Z$ ($F_{\text{P-C}} \approx 700$), and the other parameters are the same as that in Figure 2. The $F_P$ of hybrid system is relatively larger than that with $Z = 0$ due to the direct contribution from the WGM (Figures 2c and 5a). Figure 5b shows the peak value $F^p_{P-tot}$ as a function of the WGM-resonance position. The $F^p_{P-tot}$ may also exceed the sum of an individual antenna and cavity ($F_{P-A} + F_{P-C}$). As discussed above, the frequency shift plays an important role in $F^p_{P-tot}$. Thus, if we compare $F^p_{P-tot}$ with the sum of an individual cavity and a redshifted antenna ($F_{P-rA} + F_{P-C}$), the $F^p_{P-tot}$ will still not exceed the later one as shown in Figure 5b. Similarly, $F^p_{P-tot}$ cannot be relatively enlarged by increasing the antenna-cavity coupling strength ($X$) (Figure 5c). The reason is similar to that in the previously discussed case of $Z = 0$ (Figure 4c). Similarly, By increasing the $F_{P-C}$ ($Z$), the $F^p_{P-tot}$ increases while it can still not exceed ($F_{P-rA} + F_{P-C}$) as expected (Figure 5c). Note that in Figure 5c, both $X$ and $Z$ increase simultaneously with $|E/E_0|_c$. This is in consistent with a realistic situation by narrowing the gap between cavity and antenna. The antenna-cavity coupling strength can also be turned by varying the polarizability of antenna $\alpha_a$ (Figure 5d). The coupling strength ($|X^2 \varepsilon_0^2 \alpha_a \alpha_c|$) increases with $\alpha_a$, while the $F^p_{P-tot}$ still keeps no larger than ($F_{P-rA} + F_{P-C}$).

### 3.3. Cases with relatively enlarged Purcell factors

We have shown that a relatively enlarged NFE of an antenna in hybrids does not bring an enlarged $F_P$, and the limit of $F_P$ is determined by the uncoupled antenna and cavity. However, it should be noted that $F_P$ can be relatively enhanced if the near field properties of an antenna (*e.g.* $Y$) is affect by a dielectric cavity (Figure 6). The only difference between the configuration here and that of Figure 2 is that the emitter is located at the left side of the antenna, while the other parameters are the same. In this case, the NFE at L is additionally increased due to the appearance of a high-index dielectric material (Figures 6a-6c). This can be explained by the fact that the dielectric material performs a mirror-like function and induces imaginary charges [54]. Thus, the NFE at the emitter location is additionally enlarged. Note that the contributions from other modes to the baseline of NFE increase here (Figure 6a) compared to that of Figure 2, while this will not be included in the analytical model.

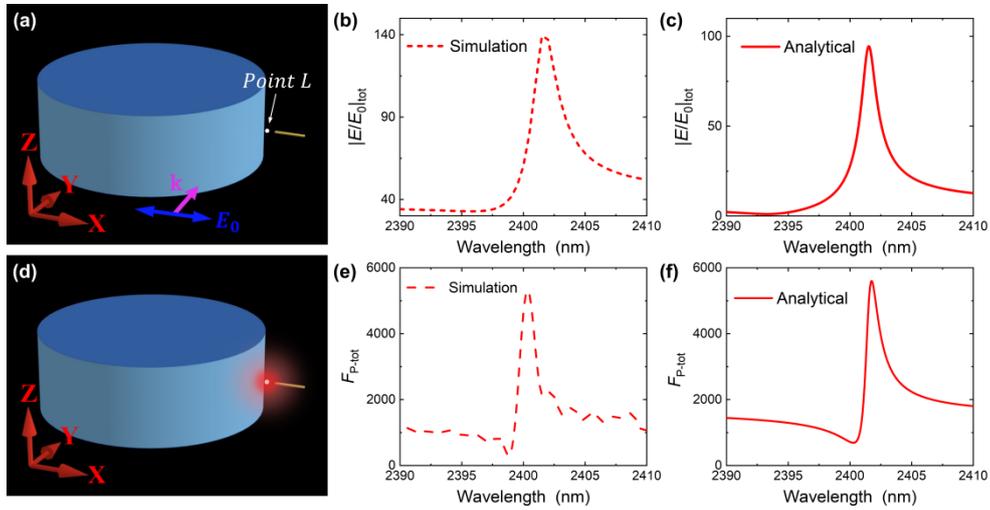

**Figure 6.** The case for a hybrid system with a modified antenna. The geometries of the disk and antenna are the same as that in Figure 2. (a) Schematic of a hybrid system excited by a plane wave. The investigated point L is between the antenna and disk, and it is 21 nm away from the rod end. (b,c) The NFE at L calculated based on direct simulations (b) and the analytical model (c). (d) Schematic of the hybrid system excited by a dipole emitter at the same point L. (e,f) $F_P$ calculated based on direct simulations (e) and the analytical model (f).

The additional enhancement of the NFE can be considered as a result of the increase in $Y$. The increase ratio can be estimated by the peak NFE responses with subtracting the effects from baselines (Figures 2b, 6b and S1). It can be obtained that the $Y$ here in the gap (Figure 6b) is increased ~1.4 times compared to the case in

Figure 2. By taken this this factor in to account, one can obtain the NFE response based on the analytical expression Eq. (1) as shown in Figure 6c. The results agree well with the direct simulations after the effects from the baselines have been subtracted. Note that the only assumption is the increased *Y*, while the other parameters used in the model are totally based on that in Figure 2.

The analytical results of $F_{\text{P-tot}}$ agree well with the simulations (Figures 6e and 6f), which again verifies the validity of the assumption of increased *Y*. It should be noted that the behavior of increased $F_{\text{P-mA}}$ remain unchanged when the disk is replaced by a semi-infinite dielectric material (Figure S4). Thus, the mirror-like effect is induced by the material not the cavity mode (WGM). It is verified that the relation between the modified $F_{\text{P-A}}$ (denoted by $F_{\text{P-mA}}$) and $|E/E_0|_{\text{mA}}$ still holds, namely, $F_{\text{P-mA}} \propto |E/E_0|^2_{mA}$ (Figure S4). Thus, the $F_{\text{P-mA}}$ is increased here compared to the unmodified one $F_{\text{P-A}}$ (Figures 2c and 6d). It can also be easily verified that the direct mode couplings between cavity and modified antenna will not relatively increase $F_{\text{P-tot}}$, namely, $F_{\text{P-tot}}$ of hybrid system can still not exceed the modified antenna $F_{\text{P-mA}}$. The reason is the same that discussed in Figure 4. On the other hand, the $F_{\text{P-tot}}$ of a hybrid system can indeed be further enhanced compared to the original components ($F_{\text{P-A}}$+ $F_{\text{P-C}}$ or $F_{\text{P-rA}}$+ $F_{\text{P-C}}$) as long as the antenna response (*e.g. Y*) is modified.

### 3.4. Hybrid systems with broad cavity modes

In the above discussion, the response of antenna in the coupled system is spectrally much broader than that of the individual WGM. Let us now consider a cavity whose response is spectrally broader than that of antenna (Figure 7a). The dielectric cavity has a magnetic dipolar mode [34] and an antenna still has a dipolar plasmon resonance (see Figure S5). The coupling between the cavity and the small plasmonic antenna now is much weaker compared to that in Figure 2. ($|X^2\varepsilon_0^2\alpha_a\alpha_c|$ << 1, see Figure S6). The direct contribution from WGM can also be ignored $Z \approx 0$ (Figure 7b). Furthermore, the plasmon and cavity modes are in the same resonant position. Thus, based on Eq. (2), the peak $|E/E_0|_{\text{tot}}$ of the hybrid system can be well reproduced by the product of the antenna $|E/E_0|_A$ and cavity $|E/E_0|_C$, namely,

$$\left|\frac{E}{E_0}\right|_{\text{tot}} \approx \left|\frac{E}{E_0}\right|_A \times \left|\frac{E}{E_0}\right|_C. \qquad (11)$$

This is numerically verified as shown in Figure 7a. As expected, the $F_{\text{P-tot}}$ of such a system is not relatively enlarged compared to that of an individual antenna (Figure 7b). This is due to the fact that the coupling in this system is weak enough. Thus, $F_{\text{P-tot}}$ of such a system reduces to ~ $F_{\text{P-A}}$. Note that $F_{\text{P-tot}}$ is slightly larger than $F_{\text{P-A}}$. This due to the small redshift of the antenna as discussed before. The additional peak around 500 nm in simulations (Figure 7b) is due to the Au material response which is ignored in our analytical model.

It should be noted that Eq. (11) does not mean the peak value of $|E/E_0|_{\text{tot}}$ reaches maximal under a weak coupling condition. This is due to the fact that the coupling strength ($|X^2\varepsilon_0^2\alpha_a\alpha_c|$) relatively increases with $|E/E_0|_C$. Now if one has two cases with two NFEs for the cavity, namely $|E/E_0|_{C-1}$ and $|E/E_0|_{C-2}$. It is also assumed that $|E/E_0|_{C-1} < |E/E_0|_{C-2}$ and $|E/E_0|_{\text{tot}-1} \approx |E/E_0|_A \times |E/E_0|_{C-1}$, where $|E/E_0|_{\text{tot}-1}$ corresponds to the NFE of first case. Even if the NFE of second

case is $|E/E_0|_{tot-1} < |E/E_0|_A \times |E/E_0|_{C-1}$, it can still be larger than $|E/E_0|_{tot-1}$. This is confirmed as shown in Figure 7c. When the coupling strength is strong enough by increasing $|E/E_0|_C$, the $|E/E_0|_{tot}$ spectrum begins to split. We can define this as a critical coupling strength. $|E/E_0|_{tot}$ reaches a maximal value at the critical coupling. The peak $|E/E_0|_{tot}$ approximately keeps this value with stronger coupling. In fact, the frequency part also plays a role in the NFE responses. A larger wavelength allows relatively larger NFE [27], namely NFE~$\sqrt{\lambda}$. This means that the relative near-field concentration ability $|E/E_0|_{tot}/\sqrt{\lambda}$ reaches maximal near the critical coupling. For stronger couplings, $|E/E_0|_{tot}/\sqrt{\lambda}$ response drops due to the splitting effect. Similar critical phenomena have been reported in other nanophtonics systems [51].

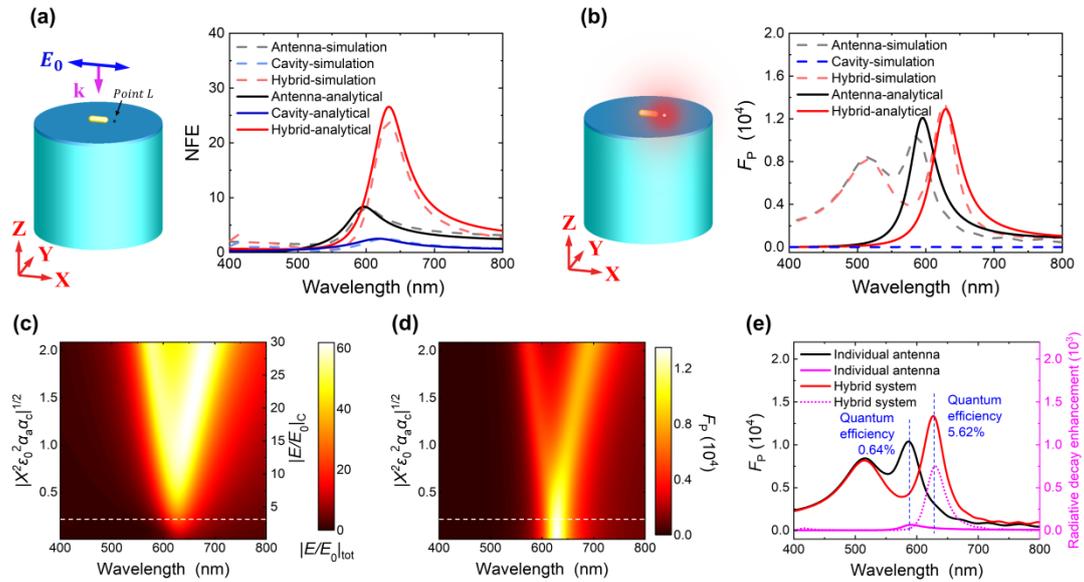

**Figure 7.** A hybrid system with a broad resonance of a dielectric cavity. The diameter and height of the disk are 160 and 160 nm, respectively. The refractive index of the disk is *n* = 3.3 (*e.g.* GaP). The length and diameter of the gold nanorod are 26 and 10 nm, respectively. The surface distance between disk and nanorod is 5 nm. The spacer between antenna and disk is a material of *n* = 1.5. (a) Left: schematic of a hybrid structure excited by a plane wave. The point L denotes a location where the NFE is investigated, and it is 2.5 nm away from the rod end. Right: The NFE at L

calculated based on direct simulations and the analytical model. The results of individual antenna and disk are also shown for comparisons. (b) Left: schematic of the same hybrid system excited by a dipole emitter. The dipole is located at the same point L. Right: $F_P$ calculated based on direct simulations and the analytical model. The results of individual antenna and disk are also shown for comparisons. (c,d) The $|E/E_0|_{tot}$ (c) and $F_{P\text{-tot}}$ (d) spectra with varying $|E/E_0|_C$ while the other parameters remain the same as before. (e) Simulated quantum efficiencies for the configuration of (d) and an individual antenna.

Figure 7d shows the calculated $F_P$ spectra with varying the coupling strength. The peak value $F_{P-tot}^p$ does not follow that of NFE as expected. The explanations are similar to that discussed before. The $F_{P-tot}^p$ can also be estimated by the relations between it and the NFE (Eqs. (7) and (8), see Figure S5). Based on Eq. (11), $F_{P-tot}^p$ under weak enough coupling (Eq. (7)) can also be expressed as

$$F_{P-tot}^p = \frac{6\pi c^2}{\omega_0^2} \frac{|E/E_0|_{tot}^2}{\sigma_{A^*}|E/E_0|_C^2}. \quad (12)$$

From the point view of a whole system, the enlarged NFE are accompanied by an enlarged effective $\sigma$ ($\sigma_{\text{eff}}$) involved in $F_P$ based on Eqs. (8) and (12). We have $\sigma_{\text{eff}} \approx \sigma_A \times |E/E_0|_C^2$ for weak couplings, and $\sigma_{\text{eff}} \approx \sigma_{\text{tot}}$ for strong couplings. From a formal perspective, the term $|E/E_0|_{tot}^2/\sigma_{\text{eff}}$ is also closely related to the well-known local density of states (LDOS) in $F_P$ [8, 9].

The coupling strength is highly related to the parameter $X$. This feature indicates that if one intends to obtain a large enough $|E/E_0|_{tot}$ with a given antenna, a proper way is to find a cavity with large enough $|E/E_0|_C$ while its extinction cross section is huge (the larger the better). Then, the coupling is weak enough and $|E/E_0|_{tot}$ will satisfy Eq. (11). Note that the $F_{P\text{-tot}}$ should still remain around that of an individual antenna $F_{P\text{-tot}} \approx F_{P\text{-A}}$. In realistic systems, the large $|E/E_0|_{tot}$ can be obtained by putting a small antenna in a large enough microcavity with an enlarged field environment. However, the $F_{\text{P-tot}}$ will not be relatively increased at all compared to an uncoupled antenna ($F_{\text{P-A}}$ or $F_{\text{P-rA}}$).

**3.5. Hybrid structures for some nanophotonic applications**

It is well known that both NFE and emission property are important for applications of enhanced Raman scattering, fluoresces and light-matter interactions. A most common situation for a hybrid structure is that a dielectric provides an enlarged-field environment for antennas. For simplicity, we shall focus on $Z \approx 0$. In terms of Raman scattering, the enhancement is usually approximated as $|E/E_0|^4$ for simplicity. This is due to the fact that both excitation and emission parts are enhanced by $|E/E_0|^2$. This works well in many plasmonic or dielectric cavities [17, 18, 23]. However, in an antenna-dielectric hybrid structure, the Raman scattering enhancement is not $|E/E_0|^4_{\text{tot}}$ anymore as the emission part is not enhanced by $|E/E_0|^2_{\text{tot}}$. As for fluorescence enhancement, the excitation part is still determined by $|E/E_0|^2_{\text{tot}}$, while the emission part is related to quantum efficiency not Purcell factor itself [15]. Pure plasmonic antennas and dielectric cavities usually surffer from low quantum efficiencies and NFEs, respectively. In this regard, an antenna-dielectric hybrid structure could work as an outstanding platform. This is because the excitation part ($|E/E_0|^2$) is easily enlarged. The quantum efficiency of a hybrid system can be largely enhanced compared to an individual plasmonic antenna [55, 56]. For example, the quantum efficiency of the above hybrid structure reaches an order of magnitude higher than that of an individual antenna (Figure 7e).

For light-matter interactions, the coupling strength between emitter and cavity can also be obtained by the $F_P$ [47, 57, 58]. Compared to an individual antenna, the effective mode volume of a hybrid system generally increases if the $F_P$ is not increased, since the effective $Q$ factor usually does not decrease. This means that simply placing a plasmon-emitter system into a dielectric cavity with enlarged-field environment cannot boost the coupling strength between an emitter and cavity, unless the property of antenna ($Y$) is modified by an antenna-dielectric configuration. Note that the relationship between $F_P$ and the emitter-cavity coupling strength is predicated on a common condition: the emitter has a spectral width that is much narrower than that of a cavity mode, and the excitation occurs in the linear region [47]. In the nonlinear region, a hybrid system would exhibit obvious differences because the nonlinear response is highly dependent on the excitation power [47].

## 4. Conclusion

In conclusion, we have theoretically investigated the NFE and $F_P$ of antenna-dielectric hybrid systems. The antenna exhibits a dipolar plasmon resonance and the cavity has Mie-like mode. Our theoretical model and direct numerical simulations agree well. The NFE around an antenna can be efficiently enlarged due to an enlarged-field environment provided by a dielectric cavity. The enlargement value is highly dependent on the $|E/E_0|_C$ and the coupling strength. The peak NFE value of a resonant hybrid system under weak coupling ($X^2 \varepsilon_0^2 \alpha_a \alpha_c \ll 1$) can reach the *product* of the antenna $|E/E_0|_A$ and cavity $|E/E_0|_C$, namely, $|E/E_0|_A \times |E/E_0|_C$.

In contrast to the NFE behavior, the $F_{P\text{-tot}}$ of a hybrid system cannot be additionally enlarged through mode couplings between antenna and cavity. This is due to the fact that the $\sigma$ response should also be considered in $F_{\text{P-tot}}$. Interestingly, the peak $F_\text{p}$ of coupled system $F_{P-tot}^p$ can be estimated by the NFE and $\sigma$ responses. For weak coupling, the $F_{P-tot}^p$ approaches that of an individual antenna ($\propto |E/E_0|_A^2/\sigma_A$). For strong enough couplings, $F_{P-tot}^p$ can be estimated by $F_\text{P} \propto |E/E_0|_{tot}^2/\sigma_{tot}$. From the view point of a whole system, the enlarged NFE are accompanied by an enlarged effective $\sigma_{\text{eff}}$ involved in $F_\text{P}$. Especially, if one takes the enlarged NFE with the original $\sigma$ for a small antenna, the $F_\text{P}$ of a hybrid cavity is largely overestimated. Furthermore, the upper limit for $F_{P-tot}^p$ of a system is also considered. The upper limit for $F_{P-tot}^p$ is set by the *sum* of the $F_\text{P}$ of individual cavity and the redshifted (or modified, if any) antenna, namely, $F_{\text{P-rA}}+ F_{\text{P-C}}$ (or $F_{\text{P-mA}}+ F_{\text{P-C}}$). Note that the $F_{\text{P-mA}}$ of a modified antenna can be larger than an individual one $F_{\text{P-rA}}$ due to the mirror-like effect brought by a dielectric structure.

    The increase in the $F_\text{P}$ is not as easy as that of the NFE in hybrid systems. This will bring different impacts on relevant nanophotonic applications. For Raman scattering, the enhancement value is smaller than $|E/E_0|_{\text{tot}}^4$. The relatively low $F_\text{P}$ shall not directly affect the fluorescence enhancement. The readily enlarged NFE and quantum efficiency would make antenna-dielectric hybrid cavities outstanding platforms for fluoresce enhancements. As for light-matter interaction, the emitter-cavity coupling strength cannot be boosted by a dielectric cavity that only provides an enlarged-field environment. This situation may improve if the $Y$ of the antenna is modified. Note that we have restricted our system with a dipolar plasmon

antenna. If the mode of an antenna involved in the near-field couplings goes beyond the dipolar approximation. The $|E/E_0|_A$ and corresponding $Y$ may be modified correspondingly. A more comprehensive understanding of this phenomenon may require more detailed investigations in the future.

## 5. Numerical Method

**Numerical simulations.** The cross-section spectra, near-field distributions, and Purcell factors can be obtained directly by a commercial finite difference time domain (FDTD) method (Ansys Lumerical). The excitation source is a total-field scattered-field plane wave for NFE studies. While for $F_P$ studies, the excitation source is an electric dipole. To improve the spatial resolution, the mesh size near the antenna region in the WGM cases is $(\Delta x, \Delta y, \Delta z) = (15\ \text{nm}, 1\ \text{nm}, 1\ \text{nm})$, while it is $(\Delta x, \Delta y, \Delta z) = (0.5\ \text{nm}, 0.5\ \text{nm}, 0.5\ \text{nm})$ for the magnetic dipole resonance case. The refractive index of the Si disk is $n = 3.5$. The surrounding index is $n = 1$ for simulations. The refractive index of Au is taken from Palik's book [59]. Perfectly matched layers were set in the $x$, $y$, and $z$ directions.

**Multipole decomposition method (MDM).** The contribution from different multipole modes to the scattering of Si disk can be calculated using the spherical multiple expansion mothed [48]. This method has been used in many micro-nano photonic structures. We can write the spherical analytic expressions with multipolar order $l$ as:

Spherical electric multipole coefficients

$$a_E(l,m) = \frac{(-i)^{l+1}kr}{h_l^{(1)}(kr)E_0[\pi(2l+1)l(l+1)]^{1/2}} \int_0^{2\pi}\int_0^{\pi} Y_{lm}^*(\theta,\phi)\hat{\mathbf{r}}\cdot\mathbf{E}_S(\mathbf{r})\sin\theta\,\mathrm{d}\theta\,\mathrm{d}\phi. \qquad (13)$$

Spherical magnetic multipole coefficients

$$a_M(l,m) = \frac{(-i)^l \eta kr}{h_l^{(1)}(kr)E_0[\pi(2l+1)l(l+1)]^{1/2}} \int_0^{2\pi} \int_0^{\pi} Y_{lm}^*(\theta,\phi) \hat{\mathbf{r}} \cdot \mathbf{H}_S(\mathbf{r}) \sin\theta \, d\theta \, d\phi. \quad (14)$$

The scattering cross-section can be calculated as

$$C_S = \frac{\pi}{k^2} \sum_{l=1}^{\infty} \sum_{m=-l}^{l} (2l+1)[|a_E(l,m)|^2 + |a_M(l,m)|^2], \quad (15)$$

where $Y_{lm}$ and $h_l^{(1)}$ are the scalar spherical harmonics and the spherical Hankel functions of first kind, respectively. $\eta$ is the impedance of the surrounding dielectric. The scattered electric field $\mathbf{E}_s$ and magnetic field $\mathbf{H}_s$ can be obtained by the above FDTD method. Furthermore, the scattering cross-section with each spherical multipole order $l$ can be obtained. It can be verified that the total scatterings calculated based on this method agree well with that from direct simulations (see Figure S7).

**Acknowledgement**

This paper was supported by the National Natural Science Foundation of China (No. 11704416), the Hunan Provincial Natural Science Foundation of China (No. 2021JJ20076).

**Conflict of Interest**

The authors declare no conflict of interest.

# Supporting Information for

# "Relations between near-field enhancements and Purcell factors in hybrid nanostructures of plasmonic antennas and dielectric cavities"

Xu-Tao Tang, Lin Ma, Yue You, Xiao-Jing Du, Hua Qiu, Xi-Hua Guan, Jun He, and Zhong-Jian Yang*

*Hunan Key Laboratory of Nanophotonics and Devices, School of Physics, Central South University, Changsha 410083, China*

*E-mail: zjyang@csu.edu.cn

## 1. Optical responses for an individual Si disk of WGM and plasmonic antenna

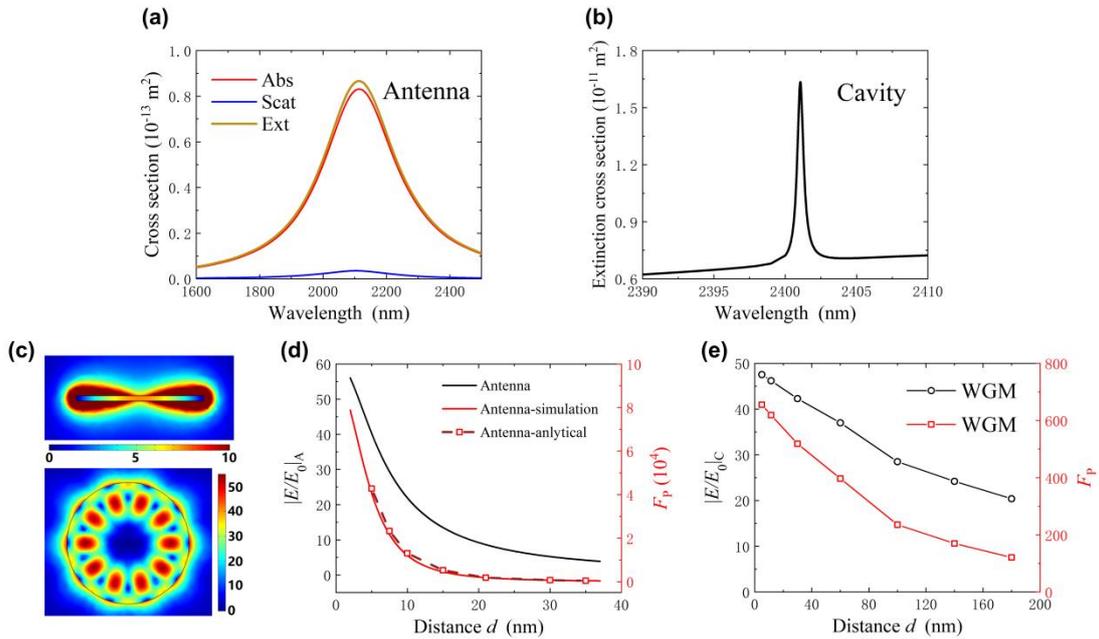

**Figure S1.** Optical responses of individual antenna and disk. The parameters are the same as that in Figure 2. (a) Extinction, scattering and absorption cross section spectra of the individual antenna. (b) Scattering and absorption cross section of the individual disk. (c) Resonant NFE distributions around the antenna (top) and disk (bottom). (d) The peak $F_P$ obtained by Eq. (7) (dashed) and direct simulations (red solid) for an antenna with different distances between emitter

and antenna end. The NFE responses are also shown (black). (e) The peak $F_P$ obtained by Eq. (7) (red) for a disk with different distances between emitter and disk surface. The NFE responses are also shown (black). For the same NFE, the $F_P$ for the two cases are quite different due to their huge differing extinction cross sections (d and e).

## 2. Relations between NFE and $F_P$ with a different $\alpha_a$

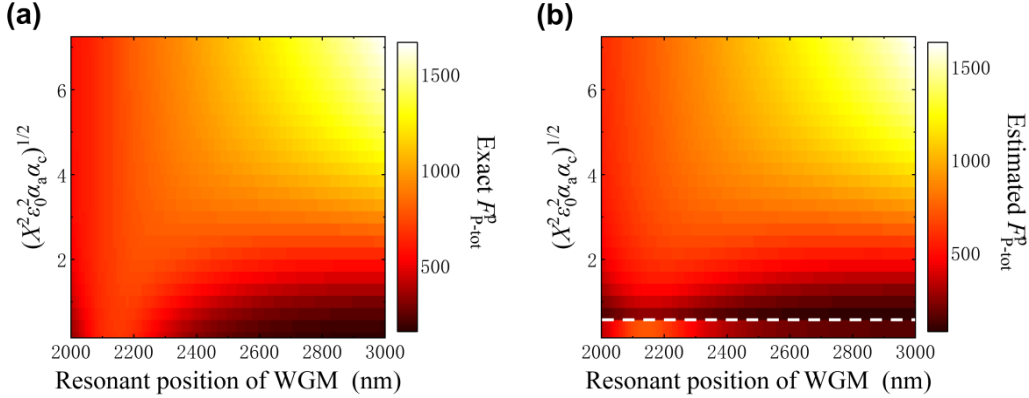

**Figure S2.** Relations between NFE and $F_P$ of hybrid structures with a different $\alpha_a$. Compared to the situation in Figure 3, the $\alpha_a$ here is increased by 2.5 times while the other parameters remain the same. The other parameters are the same as that in Figure 3. The exact (a) and estimated (b) $F_{P-tot}^p$ as a function of $|X^2\varepsilon_0^2\alpha_a\alpha_c|^{1/2}$ and the resonant position of WGM. The results above and below the dotted line are based on Eq. (8) and (7), respectively.

## 3. Limits for Purcell factor of a hybrid system with varying couplings

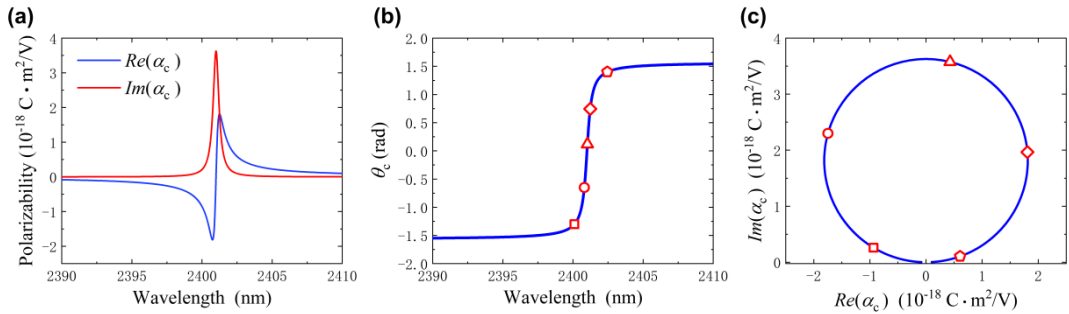

**Figure S3.** (a) The $\alpha_c$ responses spectra. This can be obtained by the MDM. (b) The corresponding $\theta_c$ as a function of wavelength in Eq. (S1). (c) The $\alpha_c$ in the complex plane. The dots marked in (b) and (c) correspond to each other.

In order to find the maximal value of the coupling part for $F_P$, namely $\text{Im}\left(\frac{\alpha_a}{1-X^2\varepsilon_0^2\alpha_a\alpha_c}\right)$, we shall treat the $\alpha_c$ first. Here, it is assumed that within the resonance range of cavity, $\alpha_a$ is taken as a constant. $X$ satisfies $\text{Im}(X^2) \approx 0$. The response of the cavity mode can be approximately taken as a Lorentz lineshape (Figure S3a). Then, $\alpha_c$ can be expressed as

$$\begin{cases} \text{Re}(\alpha_c) = D_c \sin\theta_c \cdot \cos\theta_c \\ \text{Im}(\alpha_c) = D_c \cos^2\theta_c \end{cases}, \qquad (S1)$$

where $\theta_c$ $\left(-\frac{\pi}{2} < \theta_c < \frac{\pi}{2}\right)$ is a parameter that is related to wavelength (Figure S3b) and $D_c$ is the maximal value of $\text{Im}(\alpha_c)$. The $\alpha_c$ near its resonance corresponds to a circle in the complex plane, with the center on the imaginary-axis and the circle passing through the origin (Figure S3c). This is rational for a Lorentz lineshape of $\alpha_c$. For example, at the resonance point, the imaginary part of $\alpha_c$ will reach the maximum, and the real part of $\alpha_c$ will be 0. As the imaginary part decreases, the real parts of $\alpha_c$ on the two sides of resonance point have opposite signs. We have $D_c=\text{Max}[\text{Im}(\alpha_c)]$ for the diameter of the circle. The center point of the circle is $i\frac{1}{2}D_c$.

Now let us turn to the peak value of $\text{Im}\left(\frac{\alpha_a}{1-X^2\varepsilon_0^2\alpha_a\alpha_c}\right)$. As the parameters $\alpha_a$ and $X$ are constants, $\frac{\alpha_a}{1-X^2\varepsilon_0^2\alpha_a\alpha_c}$ is still a circle in the complex plane. While the center point of the new circle is moved to $\frac{\alpha_a + X^2\varepsilon_0^2|\alpha_a|^2 i\frac{1}{2}D_c}{X^2\varepsilon_0^2\text{Im}(\alpha_a)\cdot D_c+1}$, and the radius is $\frac{X^2\varepsilon_0^2|\alpha_a|^2\cdot\frac{1}{2}D_c}{X^2\varepsilon_0^2\text{Im}(\alpha_a)\cdot D_c+1}$. The peak value of $\text{Im}\left(\frac{\alpha_a}{1-X^2\varepsilon_0^2\alpha_a\alpha_c}\right)$ corresponds to the maximal value of the $\text{Im}\left(\frac{\alpha_a}{1-X^2\varepsilon_0^2\alpha_a\alpha_c}\right)$, which is equal to the sum of the imaginary part of the center and the radius of the circle.

$$\text{Peak}\left[\text{Im}\left(\frac{\alpha_a}{1-X^2\varepsilon_0^2\alpha_a\alpha_c}\right)\right]$$

$$\approx \text{Im}\left(\frac{\alpha_a + X^2\varepsilon_0^2|\alpha_a|^2 i\frac{1}{2}D_c}{X^2\varepsilon_0^2\text{Im}(\alpha_a)\cdot D_c+1}\right) + \frac{X^2\varepsilon_0^2|\alpha_a|^2\cdot\frac{1}{2}D_c}{X^2\varepsilon_0^2\text{Im}(\alpha_a)\cdot D_c+1}$$

$$= \frac{\text{Im}(\alpha_a)+X^2\varepsilon_0^2|\alpha_a|^2 D_c}{X^2\varepsilon_0^2\text{Im}(\alpha_a)\cdot D_c+1}. \tag{S2}$$

The peak value of $\text{Im}\left(\frac{\alpha_a}{1-X^2\varepsilon_0^2\alpha_a\alpha_c}\right)$ can be further considered with varying the coupling strength $X$ and the resonance of cavity. The antenna polarization rate $\alpha_a$ varies with the resonance position of cavity. The response of the antenna mode can also be approximately taken as a Lorentz lineshape. Then, $\alpha_a$ can be written as $\begin{cases}\text{Re}(\alpha_a) = D_a \sin\theta_a \cdot \cos\theta_a \\ \text{Im}(\alpha_a) = D_a \cos^2\theta_a\end{cases}$, where $\theta_a\left(-\frac{\pi}{2}<\theta_a<\frac{\pi}{2}\right)$ is a parameter that is also related to wavelength and $D_a$ is the maximal value of $\text{Im}(\alpha_a)$, namely, $D_a=\text{Max}[\text{Im}(\alpha_a)]$. The $X^2\varepsilon_0^2 D_c$ is of a certain positive real number and it is written as $t = X^2\varepsilon_0^2 D_c$ for simplicity. Then, Eq. (S2) becomes

$$\frac{\text{Im}(\alpha_a)+X^2\varepsilon_0^2|\alpha_a|^2 D_c}{X^2\varepsilon_0^2\text{Im}(\alpha_a)D_c+1} = \frac{D_a\cos^2\theta_a+tD_a^2\cos^2\theta_a}{tD_a\cos^2\theta_a+1}. \tag{S3}$$

The derivative function of the above expression with respect to $\theta_a$ is

$$\frac{\partial}{\partial\theta_a}\left(\frac{D_a\cos^2\theta_a+tD_a^2\cos^2\theta_a}{tD_a\cos^2\theta_a+1}\right) = \frac{-D_a(tD_a+1)\sin(2\theta_a)}{(tD_a\cos^2\theta_a+1)^2}. \tag{S4}$$

It can be obtained that the right part of Eq. (S4) reaches a maximal value at $\theta_a = 0$. Then one obtains

$$\text{Max}\left[\frac{D_a\cos^2\theta_a+tD_a^2\cos^2\theta_a}{tD_a\cos^2\theta_a+1}\right] = \text{Max}[\text{Im}(\alpha_a)]. \tag{S5}$$

It can be found that the maximum value is independent of $t$, and is determined by the property of antenna. For a general $\theta_a$, it can be easily seen that $\frac{\text{Im}(\alpha_a)+X^2\varepsilon_0^2|\alpha_a|^2 D_c}{X^2\varepsilon_0^2\text{Im}(\alpha_a)D_c+1}$ increases with $t$. while it can cannot exceed the limit value $\text{Max}[\text{Im}(\alpha_a)]$. Eq. (S5) also indicates that when the narrow cavity is spectrally near the antenna resonance ($\theta_a = 0$), the coupling part of $F_P$ reaches the maximum value $\text{Max}[\text{Im}(\alpha_a)]$.

## 4. Mirror-like effects with different gaps

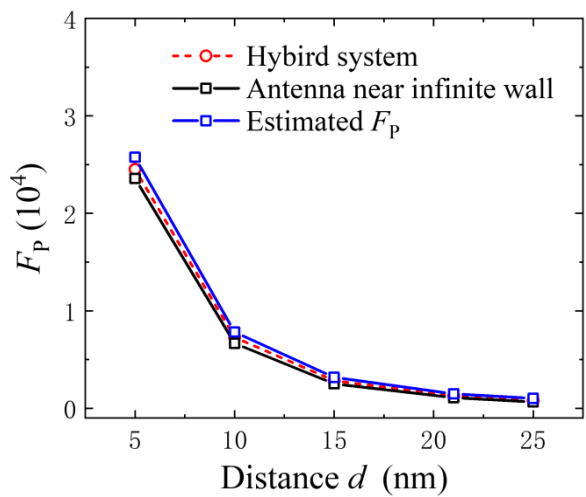

**Figure S4.** The $F_P$ responses with different gaps. The system is based on the one in Figure 6. The $F_P$ results for the case where the disk is replaced by a semi-infinite dielectric material are also shown for comparison. The $F_P$ estimated by NFEs ($|E/E_0|_{mA}$) are presented as well.

## 5. The case of a disk with a magnetic dipole response

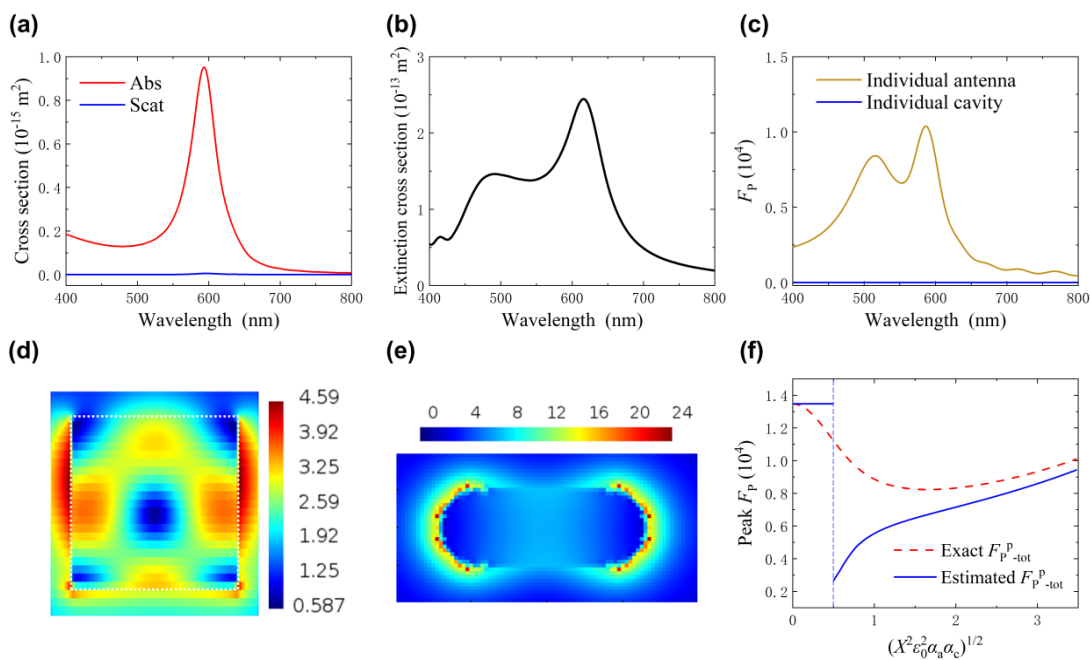

**Figure S5.** The parameters are the same as that in Figure 7. (a) Scattering and absorption cross section spectra of the individual antenna. (b) Scattering and absorption cross section of the individual disk. (c) $F_P$ spectra of individual antenna and disk at L. (d,e) Resonant NFE

distributions around the individual disk (d) and antenna (e). (f) The exact and estimated $F^{\text{p}}_{\text{P-tot}}$ with different $|X^2\varepsilon_0^2\alpha_a\alpha_c|^{1/2}$ ($|E/E_0|_c$).

## 6. The coupling parameter $|X^2\varepsilon_0^2\alpha_a\alpha_c|$ of different systems

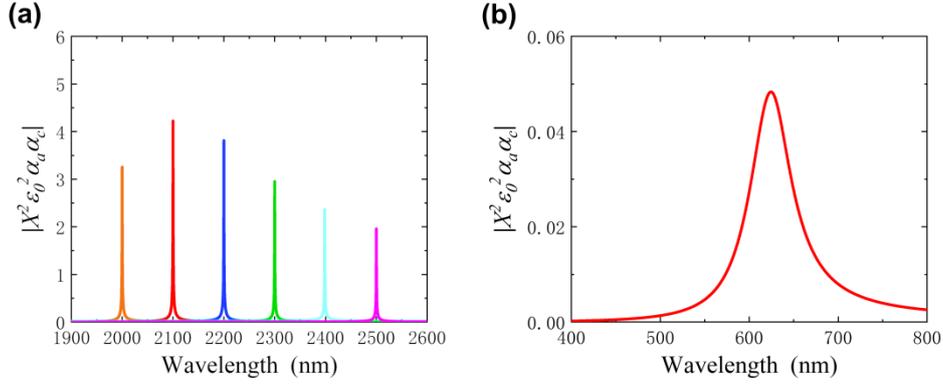

**Figure S6.** The $|X^2\varepsilon_0^2\alpha_a\alpha_c|$ response for the cases in Figures 2 and 7 are shown in (a) and (b), respectively. For the case in Figure 2, the results with WGMs of different resonant positions are also shown.

## 7. Multipole moments of a cavity calculated by MDM

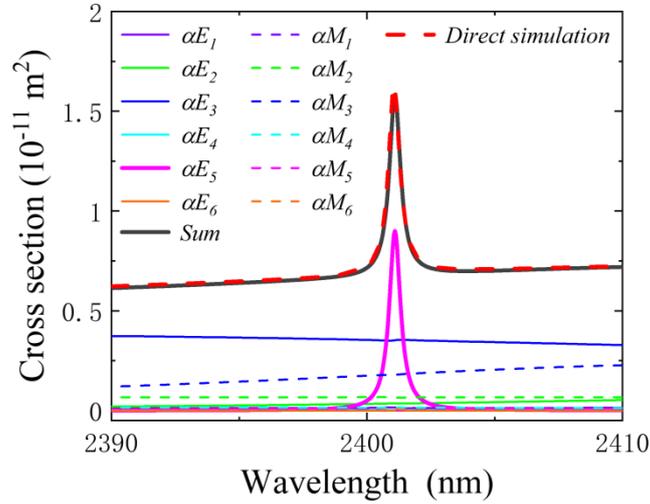

**Figure S7.** Scattering spectrum of an individual disk and the corresponding contributions from multipole moments. The disk is the same as that in Figure 2. The multipole moments are calculated by the MDM. The aE$_n$ (or aM$_n$, $n = 1,2,3…$) term corresponds to the multipole order $n$ of a$_E(n, m)$ [or a$_M(n, m)$] in the multipole coefficients [Eqs. (13) and (14)], where the contributions from all of the $m$ have been added together for each $n$.